\input harvmac

\input amssym
\input epsf

\def\unit{\relax{\rm 1\kern-.26em I}}
\def\nada{\relax{\rm 0\kern-.30em l}}
\def\tilde{\widetilde}



\def\det{{\rm det}}

\noblackbox
\def\IL{\relax{\rm I\kern-.18em L}}
\def\IH{\relax{\rm I\kern-.18em H}}
\def\IR{\relax{\rm I\kern-.18em R}}
\def\IC{\relax\hbox{$\inbar\kern-.3em{\rm C}$}}
\def\IZ{\relax\ifmmode\mathchoice
{\hbox{\cmss Z\kern-.4em Z}}{\hbox{\cmss Z\kern-.4em Z}}
{\lower.9pt\hbox{\cmsss Z\kern-.4em Z}} {\lower1.2pt\hbox{\cmsss
Z\kern-.4em Z}}\else{\cmss Z\kern-.4em Z}\fi}
\def\CM {{\cal M}}

\def\CF {{\cal F}}

\def\CO {{\cal O}}

\def\CM {{\cal M}}

\def\CO {{\cal O}}

\def\det{{\rm det}}
\def\Tr{{\rm Tr}}

\font\manual=manfnt \def\dbend{\lower3.5pt\hbox{\manual\char127}}

\def\IZ{\relax\ifmmode\mathchoice
{\hbox{\cmss Z\kern-.4em Z}}{\hbox{\cmss Z\kern-.4em Z}}
{\lower.9pt\hbox{\cmsss Z\kern-.4em Z}} {\lower1.2pt\hbox{\cmsss
Z\kern-.4em Z}}\else{\cmss Z\kern-.4em Z}\fi}
\def\half {{1\over 2}}

\def\lfm#1{\medskip\noindent\item{#1}}

\def\rt2{\sqrt{2}}
\def\irt2{{1\over\sqrt{2}}}

\def\hat{\widehat}
\def\slashchar#1{\setbox0=\hbox{$#1$}           
   \dimen0=\wd0                                 
   \setbox1=\hbox{/} \dimen1=\wd1               
   \ifdim\dimen0>\dimen1                        
      \rlap{\hbox to \dimen0{\hfil/\hfil}}      
      #1                                        
   \else                                        
      \rlap{\hbox to \dimen1{\hfil$#1$\hfil}}   
      /                                         
   \fi}

\def\foursqr#1#2{{\vcenter{\vbox{
    \hrule height.#2pt
    \hbox{\vrule width.#2pt height#1pt \kern#1pt
    \vrule width.#2pt}
    \hrule height.#2pt
    \hrule height.#2pt
    \hbox{\vrule width.#2pt height#1pt \kern#1pt
    \vrule width.#2pt}
    \hrule height.#2pt
        \hrule height.#2pt
    \hbox{\vrule width.#2pt height#1pt \kern#1pt
    \vrule width.#2pt}
    \hrule height.#2pt
        \hrule height.#2pt
    \hbox{\vrule width.#2pt height#1pt \kern#1pt
    \vrule width.#2pt}
    \hrule height.#2pt}}}}
\def\psqr#1#2{{\vcenter{\vbox{\hrule height.#2pt
    \hbox{\vrule width.#2pt height#1pt \kern#1pt
    \vrule width.#2pt}
    \hrule height.#2pt \hrule height.#2pt
    \hbox{\vrule width.#2pt height#1pt \kern#1pt
    \vrule width.#2pt}
    \hrule height.#2pt}}}}
\def\sqr#1#2{{\vcenter{\vbox{\hrule height.#2pt
    \hbox{\vrule width.#2pt height#1pt \kern#1pt
    \vrule width.#2pt}
    \hrule height.#2pt}}}}

\lref\ORaifeartaighPR{
  L.~O'Raifeartaigh,
  ``Spontaneous Symmetry Breaking For Chiral Scalar Superfields,''
  Nucl.\ Phys.\ B {\bf 96}, 331 (1975).
}

\lref\DineGM{
  M.~Dine, J.~L.~Feng and E.~Silverstein,
  ``Retrofitting O'Raifeartaigh models with dynamical scales,''
  Phys.\ Rev.\ D {\bf 74}, 095012 (2006)
  [arXiv:hep-th/0608159].
}

\lref\DineXT{
  M.~Dine and J.~Mason,
  ``Gauge Mediation in Metastable Vacua,''
  arXiv:hep-ph/0611312.
}

\lref\KitanoXG{
  R.~Kitano, H.~Ooguri and Y.~Ookouchi,
  ``Direct mediation of meta-stable supersymmetry breaking,''
  arXiv:hep-ph/0612139.
}

\lref\MurayamaYF{
  H.~Murayama and Y.~Nomura,
  ``Gauge mediation simplified,''
  arXiv:hep-ph/0612186.
}

\lref\CsakiWI{
  C.~Csaki, Y.~Shirman and J.~Terning,
  ``A simple model of low-scale direct gauge mediation,''
  arXiv:hep-ph/0612241.
}

\lref\AharonyMY{
 O.~Aharony and N.~Seiberg,
 ``Naturalized and simplified gauge mediation,''
 arXiv:hep-ph/0612308.
}

\lref\AbelUQ{
  S.~A.~Abel and V.~V.~Khoze,
  ``Metastable SUSY breaking within the standard model,''
  arXiv:hep-ph/0701069.
}

\lref\AmaritiQU{
  A.~Amariti, L.~Girardello and A.~Mariotti,
  ``On meta-stable SQCD with adjoint matter and gauge mediation,''
  arXiv:hep-th/0701121.
}

\lref\MurayamaFE{
  H.~Murayama and Y.~Nomura,
  ``Simple scheme for gauge mediation,''
  arXiv:hep-ph/0701231.
}

\lref\ISS{
  K.~Intriligator, N.~Seiberg and D.~Shih,
  ``Dynamical SUSY breaking in meta-stable vacua,''
  JHEP {\bf 0604}, 021 (2006)
  [arXiv:hep-th/0602239].
}

\lref\NelsonNF{
  A.~E.~Nelson and N.~Seiberg,
  ``R symmetry breaking versus supersymmetry breaking,''
  Nucl.\ Phys.\ B {\bf 416}, 46 (1994)
  [arXiv:hep-ph/9309299].
}

\lref\IntriligatorCP{
  K.~Intriligator and N.~Seiberg,
  ``Lectures on supersymmetry breaking,''
  arXiv:hep-ph/0702069.
}

\lref\WittenIH{
  E.~Witten,
  ``Mass Hierarchies In Supersymmetric Theories,''
  Phys.\ Lett.\ B {\bf 105}, 267 (1981).
}

\lref\NappiHM{
  C.~R.~Nappi and B.~A.~Ovrut,
  ``Supersymmetric Extension Of The SU(3) X SU(2) X U(1) Model,''
  Phys.\ Lett.\  B {\bf 113}, 175 (1982).
}

\lref\ISSii{
  K.~Intriligator, N.~Seiberg and D.~Shih,
  ``Supersymmetry breaking, R-symmetry breaking and meta-stable
  vacua," to appear.
}

\lref\RayWK{
  S.~Ray,
  ``Some properties of meta-stable supersymmetry-breaking vacua in Wess-Zumino
  Phys.\ Lett.\  B {\bf 642}, 137 (2006)
  [arXiv:hep-th/0607172].
}

\lref\ClaudsonYE{
  M.~Claudson and M.~B.~Wise,
  ``Effective Potentials For A Class Of Supersymmetric Theories,''
  Phys.\ Lett.\  B {\bf 113}, 31 (1982).
}

\lref\DineGU{
  M.~Dine and W.~Fischler,
  ``A Phenomenological Model Of Particle Physics Based On Supersymmetry,''
  Phys.\ Lett.\  B {\bf 110}, 227 (1982).
}

\lref\DineZB{
  M.~Dine and W.~Fischler,
  ``A Supersymmetric Gut,''
  Nucl.\ Phys.\  B {\bf 204}, 346 (1982).
}

\lref\AlvarezGaumeWY{
  L.~Alvarez-Gaume, M.~Claudson and M.~B.~Wise,
  ``Low-Energy Supersymmetry,''
  Nucl.\ Phys.\  B {\bf 207}, 96 (1982).
}

\lref\DimopoulosGM{
  S.~Dimopoulos and S.~Raby,
  ``Geometric Hierarchy,''
  Nucl.\ Phys.\  B {\bf 219}, 479 (1983).
}

\lref\BanksMG{
  T.~Banks and V.~Kaplunovsky,
  ``Nosonomy Of An Upside Down Hierarchy Model. 1,''
  Nucl.\ Phys.\  B {\bf 211}, 529 (1983).
}
\lref\KaplunovskyYX{
  V.~Kaplunovsky,
  ``Nosonomy Of An Upside Down Hierarchy Model. 2,''
  Nucl.\ Phys.\  B {\bf 233}, 336 (1984).
}

\newbox\tmpbox\setbox\tmpbox\hbox{\abstractfont }
\Title{\vbox{\baselineskip12pt }} {\vbox{\centerline{Spontaneous
R-Symmetry Breaking}\vskip6pt\centerline{ in O'Raifeartaigh
Models}}}
\smallskip
\centerline{David Shih}
\smallskip
\bigskip
\centerline{{\it Department of Physics, Harvard University,
Cambridge, MA 02138 USA}}
\bigskip
\vskip 1cm

\noindent We study the question of whether spontaneous $U(1)_R$
breaking can occur in O'Raifeartaigh-type models of spontaneous
supersymmetry breaking. We show that in order for it to occur,
there must be a field in the theory with R-charge different from 0
or 2. We construct the simplest O'Raifeartaigh model with this
property, and we find that for a wide range of parameters, it has
a meta-stable vacuum where $U(1)_R$ is spontaneously broken. This
suggests that spontaneous $U(1)_R$ breaking actually occurs in
generic O'Raifeartaigh models.

\bigskip

\Date{March 2007}

\newsec{Introduction}

Recently, there has been a revival of interest in low-scale SUSY
model building using renormalizable, perturbative models of
spontaneous supersymmetry breaking -- i.e.\ generalizations of the
O'Raifeartaigh model \ORaifeartaighPR\ -- in the hidden sector
\refs{\DineGM\DineXT\KitanoXG\MurayamaYF\CsakiWI\AharonyMY\AbelUQ\AmaritiQU-\MurayamaFE}.
This has been motivated in part by the realization that
O'Raifeartaigh-type models can arise naturally and dynamically in
the low-energy limit of simple SUSY gauge theories such as massive
SQCD \ISS.

In all of the recent model building attempts, one common theme has
been the R-symmetry. According to \NelsonNF, this must exist in
any generic, calculable theory of spontaneous F-term supersymmetry
breaking; but at the same time, it must be broken in order to have
nonzero Majorana gaugino masses. Because the vacuum of the
simplest O'Raifeartaigh models preserves the R-symmetry (for a
recent review of this and other facts about O'Raifeartaigh models,
see e.g.\ \IntriligatorCP), the models built to date have focused
on two mechanisms for breaking the R-symmetry, both of which
involve modifying the O'Raifeartaigh model in some way. These are:
adding explicit R-symmetry violating operators to the
superpotential; or gauging a flavor symmetry and using gauge
interactions to spontaneously break the R-symmetry. Neither of
these mechanisms for R-symmetry breaking are completely free of
problems. Explicitly breaking the R-symmetry tends to restore
supersymmetry, and this can sometimes lead to tension between
having a sufficiently long-lived meta-stable vacuum and having
sufficiently large gaugino masses. (This is not always a serious
problem; see e.g.\ the recent model of \MurayamaYF.) Meanwhile,
using gauge interactions to spontaneously break R-symmetry
typically leads to an ``inverted hierarchy" \WittenIH, which is
problematic for models of low-scale SUSY breaking. One can achieve
spontaneous R-symmetry breaking without the inverted hierarchy
(see for instance the early model of \NappiHM, and more recently
\refs{\DineXT,\CsakiWI}), but this seems to generally require
significant fine tuning of the couplings \ISSii.

In this paper, we propose to consider a much simpler mechanism for
R-symmetry breaking, namely using the perturbative dynamics of the
O'Raifeartaigh model itself to spontaneously break $U(1)_R$.

To see how this could come about, recall that in
O'Raifeartaigh-type models, there is always a pseudo-moduli space,
i.e.\ a continuous space of supersymmetry-breaking vacua with
degenerate tree-level vacuum energies \RayWK. For simplicity, we will
focus on models with a single pseudo-modulus $X$. Expanding around
the pseudo-moduli space, the superpotential of such models can
always be put in the form
\eqn\Wintro{
W = f X + {1\over2}(M^{ij}+XN^{ij})\phi_i\phi_j+\dots
}
so that the pseudo-moduli space occurs at $\phi_i=0$, with $X$
arbitrary. Here $X$ and the $\phi_i$ are chiral superfields, and
$\dots$ denote possible cubic interactions amongst the $\phi_i$
fields. These are irrelevant for the calculation of the one-loop
Coleman-Weinberg potential, which depends only on the mass
matrices of the $\phi_i$ superfields evaluated at $\phi_i=0$
(assumed to be positive definite in a neighhborhood around $X=0$):
\eqn\VCW{
V_{eff}^{(1)} = {1\over64\pi^2}\Tr\,(-1)^F
\CM^4\log{\CM^2\over\Lambda^2}
}
The R-symmetry present in \Wintro\ implies that $R(X)=2$, and
guarantees that to leading order around $X=0$, the effective
potential takes the form
\eqn\VCWexpand{
V_{eff}^{(1)} = V_0 + m_X^2|X|^2+ \CO(|X|^4)
}
The sign of $m_X^2$ then determines whether or not the R-symmetry
is spontaneously broken.

In section 2, we derive a general formula for $m_X^2$ in terms of
the matrices $M$ and $N$ appearing in \Wintro. We observe that
$m_X^2$ can in general have either sign, and we show that a {\it
necessary} condition for $m_X^2$ to be negative is that there
exists a field $\phi_i$ in \Wintro\ with R-charge other than 0 or
2.

In section 3, we construct the simplest O'Raifeartaigh model with
this property: a model with chiral fields $\phi_{1,2,3}$ having
R-charges $-1$, $1$ and $3$ respectively, and superpotential
\eqn\WORRbintro{
W = \lambda X \phi_{1}\phi_2 + m_1\phi_{1}\phi_3+\half m_2\phi_2^2
+ f X
}
We show that for a wide range of parameters, $m_X^2<0$ and there
is a local minimum of the potential with $X\ne 0$ and
spontaneously broken R-symmetry. Interestingly, this model has
runaway behavior at large fields, so the vacuum we find at $X\ne
0$ is only meta-stable. However, it can be made parametrically
long-lived in the limit $y\equiv |\lambda f/m_1 m_2|\to 0$.

Finally, in the appendix, we provide a few consistency checks of
our general formula for $m_X^2$. We apply our general formula to a
few well-known examples which we believe are representative of the
O'Raifeartaigh models used so far in model building. This includes
the original O'Raifeartaigh model \ORaifeartaighPR,
\eqn\WORorig{
W = f X + m\phi_1\phi_2+{1\over2}h X\phi_1^2
}
as well as Witten's $SU(5)$ ``inverted hierarchy" model \WittenIH,
and the simplest version of the ``rank condition" models of \ISS.
In all of these models, all of the fields have either $R=0$ or
$R=2$, and consequently, the R-symmetry remains unbroken.

The fact that spontaneous R-symmetry breaking already occurs (for
some range of the couplings) in the simplest model with more
general R-charge assignments suggests that it actually occurs
(again, for some range of the couplings) in a {\it generic}
O'Raifeartaigh-type model. One reason this may have gone unnoticed
until now is that many, if not all, of the O'Raifeartaigh models
considered to date (such as the ones in the appendix) share the
highly non-generic R-charge assignments ($R=0$ or $R=2$) of the
original model \WORorig.

The possibility of spontaneous R-symmetry breaking in
O'Raifeartaigh models opens up many new directions which would be
interesting to explore. For instance, it would be useful to find
more examples of O'Raifeartaigh models with spontaneous R-symmetry
breaking, especially ones with larger global symmetries. One
interesting question is whether the runaway behavior seen in the
example \WORRbintro\ is a general feature of these examples. Also,
it would be interesting to explore the ``retro-fitting" of these
models along the lines of \DineGM, and to search for simple,
asymptotically-free UV completions along the lines of \ISS.
Finally, the application of these ideas to phenomenology is a
promising direction which could potentially lead to new models of
supersymmetry breaking, especially models of low-scale direct
mediation.

\newsec{General results for O'Raifeartaigh models}

\subsec{A general formula for $m_X^2$}

Let us start by defining our class of models \Wintro\ more
precisely. As described in the introduction, we will consider the
most general O'Raifeartaigh type model with a single
pseudo-modulus $X$ and an R symmetry. Thus, we have a
renormalizable Wess-Zumino model consisting of a chiral superfield
$X$ and $n$ chiral superfields $\phi_i$, with canonical K\"ahler
potential and superpotential
\eqn\Wmodel{
W = f X + {1\over2} (M^{ij}+XN^{ij})\phi_i\phi_j
}
Here $M$ and $N$ are symmetric complex matrices, and we will
assume that $\det M\ne 0$. We will take $f$ to be real and
positive without loss of generality. The R-symmetry implies that
$R(X)=2$, and it constrains the form of $M$ and $N$:
\eqn\MNrel{
M^{ij}\ne 0 \Rightarrow R(\phi_i)+R(\phi_j)=2\,;\qquad N^{ij}\ne
0\Rightarrow R(\phi_i)+R(\phi_j)=0
}
Combining this with $\det M\ne 0$, we see that, in a basis where
fields of the same R-charge are grouped together, $M$ must have
the block form
\eqn\Mblock{
M = \pmatrix{  &  &  & & &  M_1\cr
              &  & & & M_2 &  \cr
               & &  & \cdot &  & \cr
               &  & \cdot & & & \cr
               & M_2^T & & &  & \cr
              M_1^T &  & & & & }
}
where the $M_i$ are individually square, non-degenerate matrices.
One consequence of this (which we will need in the next paragraph)
is that $M^{-1}$ has the same block form as $M$ and so it also
satisfies the same relations \MNrel\ as $M$.

Another important consequence of the R-symmetry is that it implies
that supersymmetry is broken. We can prove this with a direct
computation of $\det(M+XN)$. Supersymmetry is broken if this
quantity is nonzero and independent of $X$, since then the
$\phi_i$ F-terms and the $X$ F-term are incompatible. We find:
\eqn\footeq{\eqalign{
\det(M+XN) &=\exp\left(\Tr\,\log({\bf 1}_n+XM^{-1}N)\right) \det
M\cr &= \exp\left(-\sum_{k\ge 1}{(-X)^k\over
k}\Tr\,(M^{-1}N)^k\right)\det M\cr
}}
By the R-symmetry relations \MNrel, $(M^{-1}N)_i\ ^j$ is nonzero
only if $R(\phi_i)-R(\phi_j)=2$. Thus any power of $M^{-1}N$ will
have vanishing diagonal entries, and so the traces in \footeq\ all
vanish. Therefore,
\eqn\footeqii{
 \det(M+XN)=\det M\ne 0
 }
and supersymmetry is broken. Note that this is a stronger
statement than that of \NelsonNF, which argues that an R-symmetry
is only a {\it necessary} condition for spontaneous F-term
supersymmetry breaking in a generic WZ model. Here we have seen
that for O'Raifeartaigh models \Wmodel, the R-symmetry is also a
{\it sufficient} condition for supersymmetry breaking. The point
is that the models \Wmodel\ are not completely generic; in
particular, all the terms are at most linear in $X$. (They can be
made ``generic" if we also impose an obvious ${\Bbb Z}_2$ symmetry
in addition to the R-symmetry.)

The scalar potential of this model has a one-dimensional space of
extrema given by
\eqn\pmsgendef{
\phi_i=0,\qquad X\,\,\,{\rm arbitrary},\qquad V_0=f^2
}
This may or may not be the absolute minimum of the tree-level
potential, depending on the details of $M$ and $N$. In particular,
there could be other, lower energy pseudo-moduli spaces with
$\phi_i\ne 0$, and there could also be runaway behavior at large
fields. However, we will assume that the couplings are such that
\pmsgendef\ is at least a local minimum of the potential in a
neighborhood around $X=0$.

The R-symmetry implies that the effective potential $V_{eff}$ on
the pseudo-moduli space has an extremum at $X=0$:
\eqn\VeffX{
V_{eff} = V_{eff}(|X|^2) = const. + m_X^2|X|^2 + \CO(|X|^4)
}
Our goal in this subsection is to derive a general formula for
$m_X^2$ in the one-loop approximation. We will do this by
expanding the usual Coleman-Weinberg formula
\eqn\CWgen{\eqalign{
V_{eff}^{(1)} & = {1\over64\pi^2}\Tr\,(-1)^F \CM^4\log {\CM^2\over
\Lambda^2} \cr
}}
to quadratic order in $X$. Here $\CM^2$ is shorthand for $\CM_B^2$
and $\CM_F^2$, the mass matrices of the scalar and fermion
components of the superfields $\phi_i$, respectively:
\eqn\mbmf{\eqalign{ &\CM_B^2=\pmatrix{W^{\dagger}_{ik}W^{kj}&W^{\dagger}_{ijk}W^k\cr
W^{ijk}W^{\dagger}_{k}&W^{ik}W^{\dagger}_{kj}} = (\hat M+X\hat
N)^2 + f \hat N\cr
 & \CM_F^2=\pmatrix{W^{\dagger}_{ik}W^{kj}&0\cr
0&W^{ik}W^{\dagger}_{kj}} =(\hat M+X\hat N)^2
 }}
where $W^i\equiv \partial W/\partial \phi_i$, etc., and we have
defined
\eqn\massmatsii{
\hat M \equiv \pmatrix{0 & M^\dagger \cr  M & 0},\qquad \hat N
\equiv \pmatrix{0 & N^\dagger \cr  N & 0}
}
Note that we are taking $X$ to be real, which suffices for
extracting $m_X^2$, according to \VeffX.

In order to expand \CWgen\ in $X$, it helps to rewrite it in the
following form:
\eqn\CWgenrew{
 V_{eff}^{(1)}
 = -{1\over32\pi^2} \Tr\,
\int_0^{\Lambda} dv\,v^5\left( {1\over v^2+\CM_B^2}-{1\over
v^2+\CM_F^2}\right)
}
Substituting \mbmf\ into \CWgenrew\ and expanding to order $X^2$,
we obtain (after an integration by parts)
\eqn\mXgen{\eqalign{
 m_X^2 & =
 {1\over16\pi^2}\Tr\,\int_0^{\Lambda}dv\, v^3 \Bigg[ {1\over
  v^2+\hat M^2+f\hat N}\left( \hat N^2 -  \half\{\hat M,\,\hat N\}{1\over v^2+\hat M^2+f\hat N} \{\hat M,\,\hat N\}\right)\cr
   & \qquad\qquad\qquad\qquad\quad - {1\over
  v^2+\hat M^2}\left( \hat N^2 -  \half\{\hat M,\,\hat N\}{1\over v^2+\hat M^2} \{\hat M,\,\hat
  N\}\right)\Bigg]
}}
We can simplify this formula by expressing it in terms of the Hermitian matrix\foot{We thank D.~Curtin and Y.~Tsai for pointing out a typo in a previous version of the paper.}
\eqn\CFdef{
\CF(v) \equiv  (v^2+\hat M^2)^{-1/2}f\hat N(v^2+\hat M^2)^{-1/2}
}
After using \MNrel\ to eliminate some of the resulting terms, we
arrive at
\eqn\mXgeniii{
 m_X^2  ={1\over16\pi^2f^2}\int_0^{\infty}dv\,v^3\,\Tr \left[
  {\CF(v)^{4}\over 1-\CF(v)^2}
  v^2
  - 2\left( {\CF(v)^2\over 1-\CF(v)^2}\hat M\right)^2\right]
}
This is our final expression for the mass-squared of $X$ around
the origin. In the appendix, we will provide some consistency
checks of \mXgeniii\ by applying it to some well-studied examples.
Some comments on this result:

\lfm{1.} Except in the simplest models, this formula generally
provides a more efficient means of computing $m_X^2$, compared
with first diagonalizing the mass matrices, computing the
Coleman-Weinberg potential, and then expanding in $X$.

\lfm{2.}  \mXgeniii\ is a difference of two non-negative
quantities,
\eqn\mxdiff{\eqalign{
  & m_X^2 =M_1^2-M_2^2\cr
  }}
where
\eqn\mxdiffii{\eqalign{
  & M_1^2 = {1\over16\pi^2f^2}\int_0^{\infty}dv\,v^5\,\Tr
  {\CF(v)^{4}\over 1-\CF(v)^2}
  \cr
   &  M_2^2 = {1\over8\pi^2f^2}\int_0^{\infty}dv\,v^3\,\Tr\left[
  \left( {\CF(v)^2\over 1-\CF(v)^2}\hat M\right)^2\right]
  }}
In the absence of any inequalities relating these quantities,
$m_X^2$ will be of indefinite sign. This raises the possibility of
spontaneous R-breaking in an O'Raifeartaigh type model, without
the need for gauge interactions. In the next subsection, we will
derive a necessary condition which must be satisfied in order for
this to occur.

\subsec{R-charge assignments and spontaneous $U(1)_R$ breaking}

We claim that in O'Raifeartaigh models where all fields have
R-charge either 0 or 2, $M_2^2=0$ and $m_X^2=M_1^2>0$. Thus, in
order for $m_X^2$ to be negative, there must be at least one field
in the theory with R-charge other than 0 or 2.\foot{Note that if
the theory has other global $U(1)$ symmetries, then any
combination of these with the R-symmetry is also an R-symmetry. So
a more precise statement of our claim is: if there is some choice
of the R-symmetry such that all the fields have $R=0$ or $R=2$,
then $m_X^2=M_1^2>0$.}

The proof is straightforward. If all of the $\phi_i$ have R-charge
either 0 or 2, then according to \MNrel, $N$ and $M$ must take the
form
\eqn\NMbd{
M=\pmatrix{ 0 & M_{02} \cr M^T_{02} & 0},\qquad N= \pmatrix{
N_{00} & 0 \cr 0 & 0}
}
in a basis where the fields with R-charge 0 and 2 are grouped
together into blocks. Hence,
\eqn\hatMhatNbd{
\hat M = \pmatrix{ 0 & 0 & 0 & M_{02}^*\cr 0 & 0 & M_{02}^\dagger
& 0\cr 0 & M_{02} & 0 & 0\cr M_{02}^T & 0 & 0 & 0},\qquad
 \hat N =  \pmatrix{
0 & 0 & N_{00}^\dagger &  0 \cr 0 & 0 & 0 & 0\cr N_{00} & 0 & 0 &
0\cr 0 & 0 & 0 & 0}
}
Substituting into \CFdef, we find that $\CF$ has the same block
form as $\hat N$:
\eqn\calFbd{
 \CF = \pmatrix{ 0 & 0 & \CF_{00}^\dagger & 0
\cr 0 & 0 & 0 & 0 \cr \CF_{00} & 0 & 0 & 0 \cr 0 & 0 & 0 & 0}
}
Finally, substituting into \mxdiffii\ gives
\eqn\mXgenstsimp{
\left( {\CF^2\over 1-\CF^2}\hat M\right)^2 =
 \pmatrix{ 0 & 0 & 0 & {\CF_{00}^\dagger\CF_{00}\over 1-\CF_{00}^\dagger\CF_{00}}M_{02}^*\cr 0 & 0 &
 0
& 0\cr 0 & {\CF_{00}\CF_{00}^\dagger\over
1-\CF_{00}\CF_{00}^\dagger}M_{02} & 0 & 0\cr 0 & 0 & 0 & 0}^2 = 0
}
so $M_2^2$ vanishes. Meanwhile,
\eqn\mXgenstsimpii{
\Tr\,{\CF^4\over 1-\CF^2} =
 2\,\Tr\,{(\CF_{00}^\dagger\CF_{00})^2\over
 1-\CF_{00}^\dagger\CF_{00}} > 0
}
so $m_X^2=M_1^2$ is non-vanishing and positive. This completes the
proof of the claim.

The class of O'Raifeartaigh models where all fields have $R=0$ or
$R=2$ may seem very non-generic, but it actually characterizes
many (if not all!) of the O'Raifeartaigh models studied in the
literature. In all of these models, the R-symmetry was found to be
unbroken in the vacuum, and now we see that this is a direct
consequence of the R-charge assignments. (An earlier hint of this
came from the work of \ClaudsonYE, who showed that the R-symmetry
is unbroken in a particular subset of models of this type.) The
examples in the appendix will illustrate this point in detail.

\newsec{The simplest O'Raifeartaigh model with spontaneous $U(1)_R$
breaking}

\subsec{The model and its vacuum}

We have seen that a necessary condition for spontaneous R-symmetry
breaking in O'Raifeartaigh models is that there is a field with
R-charge other than 0 or 2. In this section we will construct and
analyze the simplest O'Raifeartaigh model of this type. We will
see that, indeed, spontaneous R-symmetry breaking occurs in this
model for a wide range of the parameters.

Our ``simplest model" is constructed in the following way. Suppose
we have a field $\phi_1$, in addition to $X$, with $R(\phi_1)=n$.
This field must have a mass term, so there must be another field
$\phi_3$ with $R(\phi_3)=2-n$. In addition, it should feel the SUSY
breaking, so there must be a field $\phi_2$ with $R(\phi_2)=-n$ to
allow for the coupling $X\phi_1\phi_2$. Finally, $\phi_2$ needs a
mass term, so there must be a field $\phi_4$ with $R(\phi_4)=2+n$.
Thus we need at least four fields in addition to $X$ -- except in
the degenerate cases $n=0$ and $n=\pm 1$. When $n=0$, $\phi_1$ and
$\phi_2$ can be identified, as can $\phi_3$ and $\phi_4$. This is
the original O'Raifeartaigh model. In the cases $n=-1$ or $n=1$,
$\phi_4$ or $\phi_3$ are redundant, respectively. The two are
equivalent after a trivial field redefinition, so we will take
$n=-1$ without loss of generality. The most general renormalizable
superpotential consistent with the R-symmetry is:
\eqn\WORRb{
W = \lambda X \phi_1\phi_2 + m_1\phi_1\phi_3+\half m_2\phi_2^2 + f
X
}
This is the simplest O'Raifeartaigh-type model containing a field
with $R\ne 0,\,2$.

The scalar potential is
\eqn\VscORRb{
V = |\lambda \phi_1\phi_2+f|^2 + |\lambda X \phi_2+m_1\phi_3|^2 +
 |\lambda X \phi_1+m_2\phi_2|^2 + |m_1\phi_1|^2
 }
By rotating the phases of all the fields, we can always take all
the couplings to be real and positive, without loss of generality.
The extrema of the potential consist of a pseudo-moduli space
\eqn\pmsdefrb{
\phi_i=0,\qquad X\,\,\,{\rm arbitrary}
}
In addition, there is runaway behavior as $\phi_3\to\infty$
\eqn\runaway{
 X = \left({m_1^2m_2\phi_3^2\over\lambda^2 f}\right)^{1/3},\qquad
 \phi_1 = \left({f^2m_2\over \lambda^2m_1\phi_3}\right)^{1/3},\qquad \phi_2=\left({f m_1\phi_3\over
\lambda m_2}\right)^{1/3},\qquad \phi_3\to\infty
}
As a check, note that the scaling exhibited in \runaway\ is
consistent with the R-symmetry.

The runaway behavior at large fields implies that the
pseudo-moduli space is not an absolute minimum of the potential.
However, as long as
\eqn\Xbound{
|X| < {m_1\over\lambda}{1-y^2\over 2y}
}
the pseudo-moduli space is a local minimum of the potential. Here
we have defined
\eqn\ydef{
y  ={\lambda f\over m_1m_2}
}
For $|X|$ larger than the bound, a linear combination of the
$\phi_i$ fields becomes tachyonic, and the system can roll down
classically into a runaway direction.

Note that for $y>1$, the pseudo-moduli space is unstable for all
$X$, while for $y<1$ there is some neighborhood around $X=0$ which
is stable. The size of this neighborhood grows monotonically as
$y\to 0$.

Now let us compute the mass-squared of $X$ around the origin. In
this simple example, it is straightforward to compute directly the
full Coleman-Weinberg potential, and then expand around the origin
to extract $m_X^2$. However, let us compute it using the general
formula \mXgeniii, so that we can see the effect of having a field
with $R\ne 0,2$. The matrices $M$ and $N$ in \Wmodel\ are
\eqn\MNrb{
M = \pmatrix{ 0 & 0 & m_1 \cr 0 & m_2 & 0 \cr m_1 & 0 & 0},\qquad
N = \pmatrix{ 0 & \lambda & 0 \cr \lambda & 0 & 0 \cr 0 & 0 & 0}
}
Substituting into \CFdef\mXgeniii, we find that both terms of
\mXgeniii\ are now nonzero -- a consequence of the fact that there
are fields with $R\ne 0,2$ in the model. The expressions are quite
complicated, but they simplify in the small $y$ limit:
\eqn\terms{\eqalign{
 & M_1^2 = {m_1^2\lambda^2y^2\over8\pi^2} {r^2(-4r^2\log
r+r^4-1)\over (r^2-1)^3}+\CO(y^4)\cr
 & M_2^2=
{m_1^2\lambda^2y^2\over 4\pi^2}{ r^4((r^2+1)\log r-r^2+1)\over
(r^2-1)^3}+\CO(y^4)
}}
where we have defined $r=m_2/m_1$. Their difference is
\eqn\VCWORRb{\eqalign{
  m_X^2 = M_1^2-M_2^2=-{m_1^2\lambda^2 y^2\over8\pi^2}
   {r^2(2r^2(r^2+3)\log r-(3r^4-2r^2-1))\over(r^2-1)^3} + \CO(y^4)
   }}
A plot of the function of $r$ appearing in \VCWORRb\ is
shown in figure 1. We see that for small $r$, $m_X^2$ is positive,
while for large $r$ it is negative. The turnover point is at
\eqn\rturnover{
r=r_*\approx 2.11
}
This is valid in the small $y$ limit; more generally, $r_*$ is a
function of $y$.

We have shown that for $r>r_*$, the mass-squared of $X$ is
negative around the origin, but we still need to check that there
is a local minimum with $X\ne 0$. Let us check this analytically,
again in the small $y$ limit. Expanding the effective potential to
$\CO(|X|^4)$, we find
\eqn\Veffquart{
V_{eff}^{(1)} = const. + m_X^2 |X|^2 + {1\over4}\lambda_X |X|^4 +
\CO(|X|^6)
}
where
\eqn\lambdaX{
   \lambda_X = {3\lambda^4y^2\over
   8\pi^2}{r^2(12r^2(r^4+5r^2+2)\log r-19r^6-9r^4+27r^2+1)\over
   (r^2-1)^5} + \CO(y^4)
  }
For $r\ge r_*$, $m_X^2\le 0$, but $\lambda_X$ is strictly
positive. Balancing the quadratic and quartic terms, we find a
local minimum at
\eqn\Xminapprox{
|X|^2\approx {2|m_X^2|\over \lambda_X} = {m_1^2\over\lambda^2}f(r)
}

\bigskip

\centerline{\epsfxsize=0.55\hsize\epsfbox{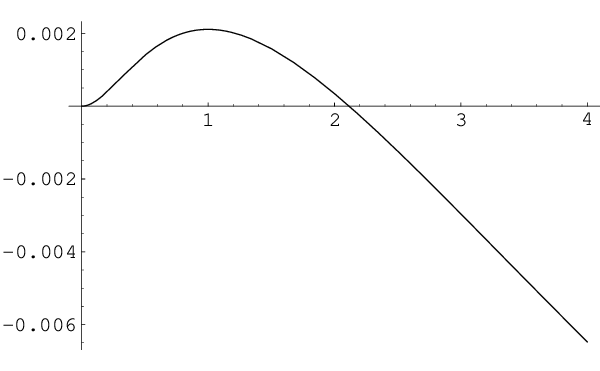}}
\noindent{\ninepoint\sl \baselineskip=8pt {\bf Figure 1}:{\sl $\;$
A plot of $m_X^2$ vs. $r=m_2/m_1$ in the small $y=\lambda f/m_1
m_2$ limit. We see that for $r> r_*\approx 2.11$, the mass-squared
of $X$ around the origin is negative, and the R-symmetry is
spontaneously broken.
}}

\noindent Since $f(r)$ is a monotonically increasing and unbounded
function of $r\ge r_*$, satisfying $f(r_*)=0$, this approximation
is valid for some range of $r$ above $r_*$, where $|X|$ is
sufficiently small that the expansion \Veffquart\ can be trusted.
Note that \Xminapprox\ implies that $|X|$ is $\CO(y^0)$ in the
$y\to 0$ limit, so by taking $y$ parametrically small, we can
always satisfy the tachyon-free bound \Xbound. We conclude that,
at least for infinitesimal $y$ and some range of $r$ above $r_*$,
and there is a local minimum of the potential at $|X|\ne 0$.

For more general values of the parameters, the analytic approach
to minimizing the potential becomes intractable. However, for a
given set of couplings $(r,\,y)$, it is straightforward to
numerically minimize the full one-loop Coleman-Weinberg potential.
One can then scan over a grid of couplings and determine the
region in coupling space where the minimum of the potential breaks
$U(1)_R$ and satisfies \Xbound. The result of such a numerical
analysis is shown in figure 2.

To summarize, we find that for a wide range of parameters there is
a local $U(1)_R$-breaking minimum of the potential at $X\ne 0$ and
satisfying \Xbound. A plot of the full one-loop effective
potential is shown in figure 3 for a representative choice of the
parameters. In general, $X\sim \CO(m_*)$ where $m_*$ is some
characteristic mass scale of the model determined by $m_1$, $m_2$
and $f$. Therefore, this is spontaneous $U(1)_R$-breaking without
inverted hierarchy. This could be useful for low-scale model
building, especially since the more conventional approach of
gauging a flavor symmetry seems to lead to a non-hierarchical
$U(1)_R$-breaking phase only in a relatively narrow window of
coupling space \ISSii.

\bigskip

\centerline{\epsfxsize=0.42\hsize\epsfbox{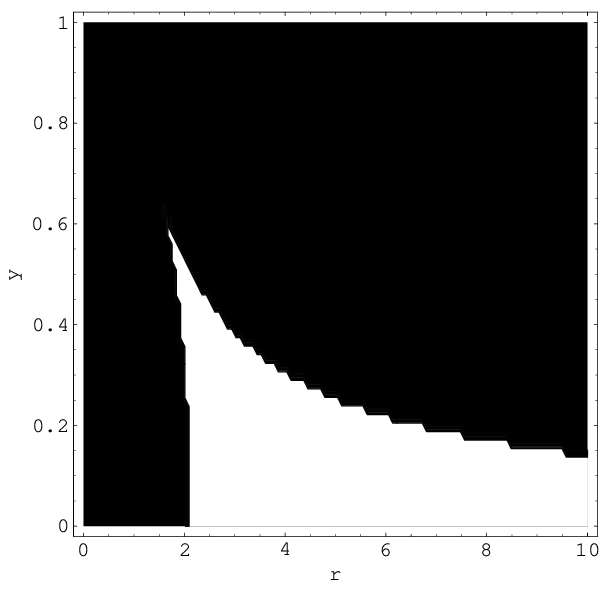}}
\noindent{\ninepoint\sl \baselineskip=8pt {\bf Figure 2}:{\sl $\;$
A plot of the region (shown in white) in the $r$, $y$ plane where
there is a $U(1)_R$-breaking local minimum of the potential
satisfying \Xbound.
}}

\bigskip

\bigskip

\centerline{\epsfxsize=0.55\hsize\epsfbox{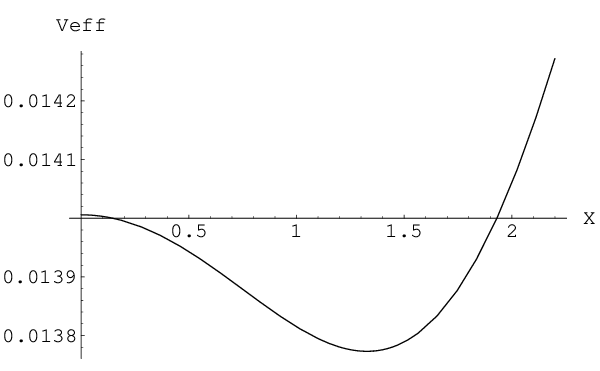}}
\noindent{\ninepoint\sl \baselineskip=8pt {\bf Figure 3}:{\sl $\;$
A plot of $V_{eff}^{(1)}$ vs.\ $|X|$ for the theory \WORRb\ with
$m_1=1$, $m_2=4$, $\lambda=1$, $y=0.2$. The R-breaking local
minimum of the potential occurs at $|X|\approx 1.3$. According to
the bound \Xbound, a transverse direction (a linear combination of
the $\phi_i$ fields) becomes tachyonic for $|X|>2.4$.
}}

\bigskip

\subsec{Lifetime estimate}

Finally, let us briefly discuss the lifetime of the R-symmetry
breaking vacuum found in the previous subsection. This vacuum is
only meta-stable, because of the runaway directions at large
fields \runaway. However, we expect that the lifetime of the
vacuum is controlled by the parameter $y$, and in the small $y$
limit it is parametrically long-lived. We can estimate the
lifetime by noting that along the runaway direction \runaway, the
value of $|X|$ at which the potential energy becomes equal to the
false vacuum energy $|f|^2$ is
\eqn\Xstar{
|X| = {m_1\over \lambda} y^{-1}
}
For smaller values of $|X|$, the potential energy is larger than
$f^2$ along the runaway direction. So this indicates that the
barrier width scales like $y^{-1}$ (since the meta-stable vacuum
is at $X\sim \CO(y^{0})$ according to \Xminapprox). Since the
barrier height is $\CO(y^0)$, this is enough to guarantee that the
meta-stable vacuum is parametrically long-lived in the $y\to 0$
limit.

\bigskip
\bigskip

\noindent {\bf Acknowledgments:}

We would like to thank Michael Dine, Ken Intriligator, Patrick
Meade and Nathan Seiberg for useful discussions and comments on
the draft. This work has been supported in part by the DOE grant
DE-FG0291ER40654.

\appendix{A}{Some examples}

In this appendix, we will apply our general formula for $m_X^2$
\mXgeniii\ to some of the most well-studied O'Raifeartaigh-type
models, which all happen to have the property that all fields have
either $R=0$ or $R=2$. This will serve as a consistency check of
the above calculations. In all of the examples, we will assume
without loss of generality that all the couplings are real and
positive.

\subsec{Example 1: the original O'Raifeartaigh model}

The first example is the basic O'Raifeartaigh model. The simplest
generalizations of this model were featured in some of the
earliest SUSY model building attempts
\refs{\DineGU\DineZB-\AlvarezGaumeWY,\NappiHM}:
\eqn\WORbasic{
W={1\over2}h X \phi_1^2+m \phi_1\phi_2+f X
}
Here $R(X)=R(\phi_2)=2$ and $R(\phi_1)=0$. When $y\equiv {h f\over
m^2} <1$, the pseudo-moduli space is \pmsgendef, with
$M=m\pmatrix{0 & 1\cr 1 & 0}$ and $N=h\pmatrix{1 & 0\cr 0 & 0
}$. From this, we find
\eqn\CNORi{
\CF = \pmatrix{ 0 & 0 & {h f\over v^2+m^2} & 0\cr 0 & 0 & 0 & 0\cr
{h f\over v^2+m^2} & 0 & 0 & 0\cr 0 & 0 & 0 & 0}
}
Substituting into \mXgeniii, we see that the second term is zero,
while the first term is nonzero, yielding
\eqn\mXORi{\eqalign{
m_X^2 &= {1\over16\pi^2f^2}\int_{0}^{\infty}dv\,{2f^4h^4 v^5 \over
(v^2+m^2)^2((v^2+m^2)^2-h^2f^2)}\cr
 & = {h^2 m^2\over 32\pi^2} y^{-1}(
 (1+y)^2\log(1+y)-(1-y)^2\log(1-y)-2y)\cr
}}
which is the expected (positive) result.

What about when $y>1$? Although the potential is minimized along a
pseudo-moduli space
\eqn\pmsygt{
 \phi _2=-{hX\over m}\phi_1,\qquad \phi_1= \pm {im\over h}\sqrt{2(y-1)},\qquad X\,\,\,{\rm arbitrary}
 }
which is not of the form \pmsgendef, the general formula
\mXgeniii\ for $m_X^2$ can still be applied after performing a
unitary transformation and constant shift on the fields.
Specifically, we take
\eqn\unitary{\eqalign{
 \phi_1 &=\tilde\phi_1 \pm {im\over h}\sqrt{2(y-1)},\qquad
 \pmatrix{\phi_2\cr X} = \pmatrix{ \cos\theta & \mp i\sin\theta \cr
 \mp i\sin\theta  &\cos\theta}\pmatrix{\tilde\phi_2\cr \tilde X}\cr
}}
with $\cos\theta={1\over \sqrt{2y-1}}$, and this yields a
superpotential of the form
\eqn\Wunitary{
W = {1\over 2}\tilde h \tilde X\tilde \phi_1^2+ \tilde m
\tilde\phi_1\tilde\phi_2 +\tilde f \tilde X + \half \tilde\lambda
\tilde\phi_1^2\tilde\phi_2
}
where
\eqn\tildeparam{
\tilde h = {h\over \sqrt{2y-1}},\qquad \tilde m =
m\sqrt{2y-1},\qquad \tilde f = f{\sqrt{2y-1}\over y},\qquad
\tilde\lambda = \mp {ih\sqrt{2(y-1)}\over\sqrt{2y-1}}
}
Aside from the extra cubic term, this model is again of the
O'Raifeartaigh form \WORbasic, and it has $\tilde y = {1\over
2y-1}<1$. The cubic term does not affect the mass matrices around
the pseudo-moduli space $\tilde\phi_i=0$, and so the calculation
of the one-loop effective potential is unchanged. Thus, the
general formula \mXgeniii\ still applies, and it yields \mXORi\
with the couplings replaced by \tildeparam.

\subsec{Example 2: Witten's $SU(5)$ ``inverted hierarchy" model}

This model has also featured in many early model building attempts
\refs{\WittenIH,\DimopoulosGM\BanksMG-\KaplunovskyYX}
\eqn\Winvh{
W = {1\over2}h X \Tr\,A^2 + \half \lambda \Tr\,A^2B + f X
}
where the original motivation was to explain dynamically the
hierarchy between the weak scale and the GUT scale. Here $A$ and
$B$ are adjoints of an $SU(5)$ global symmetry, which in the
original models was gauged and identified with the GUT group.
Supersymmetry is broken because the $X$ and $B$ equations of
motion are inconsistent \WittenIH. The R-charge assignments are
$R(X)=R(B)=2$ and $R(A)=0$.

This model has a one-dimensional pseudo-moduli space (up to global
symmetries) given by
\eqn\pmsinvh{
 A=\pm i\sqrt{2fh\over 30h^2+\lambda^2}\,{\rm
 diag}(2,2,2,-3,-3),\qquad B = {h X\over\lambda}\,{\rm
 diag}(2,2,2,-3,-3),\qquad X\,\,\,{\rm arbitrary}
 }
which spontaneously breaks $SU(5)\to SU(3)\times SU(2)\times
U(1)$. As in the $y>1$ phase of the previous example, we can put
this in the form \pmsgendef\ by expanding around this
pseudo-moduli space with a unitary transformation and constant
shift:
\eqn\pmsexpandinvh{\eqalign{
 &A = \left(\pm i\sqrt{{2fh\over 30h^2+\lambda^2}}+{a\over\sqrt{30}} \right)\pmatrix{2 & 0 \cr 0 & -3} +
 \pmatrix{ A_{33} & A_{32} \cr  A_{23} &
 A_{22}}\cr
 &B = {1\over\sqrt{30h^2+\lambda^2}}\left( h \tilde X+
 {\lambda b\over\sqrt{30}}\right) \pmatrix{2 & 0 \cr 0 & -3}
 + \pmatrix{B_{33} & B_{32} \cr B_{23} &
 B_{22}}\cr
 &X ={1\over\sqrt{30h^2+\lambda^2}}\left(
 \lambda \tilde X-\sqrt{30}h b\right)
 }}
Here $A_{22}$, $B_{22}$ and $A_{33}$, $B_{33}$ are adjoints of
$SU(2)$ and $SU(3)$ respectively; and $A_{23}$, $A_{32}$,
$B_{23}$, $B_{32}$ are bifundamentals of $SU(2)\times SU(3)$. With
this transformation, the superpotential becomes
\eqn\Wexpandinvh{\eqalign{
& W = \tilde f \tilde X  + {1\over2}\tilde h \tilde X\Tr\,(3
A_{33}^2+A_{32}A_{23}-2A_{22}^2) +\tilde m
\Tr\,(4A_{33}B_{33}-A_{32}B_{23}-A_{23}B_{32}-6A_{22}B_{22})\cr
 & \qquad +
\tilde M a b + (cubic)
}}
with
\eqn\tildeparamdef{
\tilde f = {\lambda f\over \sqrt{30h^2+\lambda^2}},\qquad \tilde h
={\lambda h \over\sqrt{30h^2+\lambda^2}},\qquad \tilde m
=\pm{i\lambda\over\sqrt{2}} \sqrt{fh\over 30h^2+\lambda^2},\qquad
\tilde M = \mp i\sqrt{2fh}
}
From this, it is straightforward to compute the matrices $M$ and
$N$, substitute into \CFdef\mXgeniii, and integrate. (Note that
although there are Goldstone bosons in the spectrum because of the
spontaneously broken $SU(5)$, the matrix $M$ is still
non-degenerate.) We again find that the second term of \mXgeniii\
is zero, while the first term is not. The final result is:
\eqn\mXsqinvh{
m_X^2 = {3f h^3 \lambda^4(968\log 11-288\log 9+600\log
5-3038\log2-529)\over 32\pi^2(30h^2+\lambda^2)^2} > 0
}

\subsec{Example 3: the $SU(2)$ ``rank condition" model}

For our final example, let us study the simplest ``rank condition"
model of \ISS, namely the one with $SU(2)$ global symmetry. The
$SU(N_f)$ generalizations of this model have featured in many of
the recent efforts at O'Raifeartaigh model building referred to in
the introduction. The superpotential is
\eqn\Wrc{
W = h\Tr\,\Phi \varphi\tilde\varphi - h\mu^2\Tr\,\Phi
}
where $\Phi$ is an ${\bf adj\oplus 1}$ and $\varphi$,
$\tilde\varphi$ are doublets under the $SU(2)$ global symmetry.
This model has an R-symmetry where $R(\Phi)=2$ and $R(\varphi)=0$.
The absolute minimum of the tree-level scalar potential occurs
along the pseudo-moduli space (modulo global symmetries)
\eqn\pmsrc{
\Phi=\pmatrix{ 0 & 0 \cr 0 & X},\qquad \varphi=\pmatrix{Y \cr 0
},\qquad \tilde\varphi=\pmatrix{{\mu^2\over Y}\cr 0},\qquad
}
Although the pseudo-moduli space $(X,Y)$ is two-dimensional in
this case, we can imagine working at fixed $Y$, and then the
theory for $X$ is still of the form \Wmodel. To see this
explicitly, we expand around a point on \pmsrc,
\eqn\pmsrcii{
\Phi=\pmatrix{ \delta\Phi_{00} & \delta\Phi_{01} \cr
\delta\Phi_{10} & X},\quad
\varphi=\pmatrix{Y+f_1(Y)\delta\chi_++f_2(Y)\delta\chi_- \cr
\delta\varphi_1},\quad \tilde\varphi=\pmatrix{{\mu^2\over
Y}+f_2^*(Y)\delta\chi_+-f_1(Y)\delta\chi_- \cr
\delta\tilde\varphi_1}
}
where $f_1(Y)=(1+|Y|^4/\mu^4)^{-1/2}$ and
$f_2(Y)=(1+|Y|^4/\mu^4)^{-1/2}Y^2/\mu^2$. Then the superpotential
becomes
\eqn\Wrcii{
W = \Big( h X(\delta\varphi_1\tilde\delta\varphi_1-\mu^2)
 + {h\mu^2\over Y}\delta\Phi_{01}\delta\varphi_1+ h Y
 \delta\Phi_{10}\delta\tilde\varphi_1 \Big) + \Bigg( {h\sqrt{\mu^4+|Y|^4}\over
 Y}\delta\Phi_{00}\delta\chi_+\Bigg) + ({\rm cubic})
}
where (cubic) refers to terms that are cubic in the fluctuations.
We see that to quadratic order, the model splits into two
disconnected sectors -- the first is an O'Raifeartaigh-type model
of the form \Wmodel\ and the second is supersymmetric and massive.
At one-loop, the effective potential for $X$ comes solely from the
first sector. It has
\eqn\MNrc{
M=\pmatrix{0 & 0 & {h\mu^2\over Y} & 0\cr 0 & 0 & 0 & h Y \cr
{h\mu^2\over Y} & 0 & 0 & 0 \cr  0 & h Y & 0 & 0\cr},\qquad
 N = \pmatrix{0 & h & 0 & 0 \cr h & 0 & 0 & 0 \cr 0 & 0 & 0 & 0\cr 0& 0&0&0}
 }
From this, we compute $\CF$ using \CFdef\ and substitute into
\mXgeniii\ to find $m_X^2$. We again find that the second term
vanishes, while the first term is nonzero and gives the positive
result
\eqn\mXrc{\eqalign{
m_X^2 &= {1\over16\pi^2h^2\mu^4}\int_{0}^{\infty}v^3\left({
4h^8\mu^8\over (v^2+h^2|Y|^2)(v^2+{h^2\mu^4\over |Y|^2})
(v^2+h^2|Y|^2+{h^2\mu^4\over |Y|^2})}\right)\cr
 &= {h^4\over 8\pi^2|Y|^2(|Y|^4-\mu^4)}\left( |Y|^8\log
 {|Y|^4+\mu^4\over |Y|^4} -\mu^8 \log {|Y|^4+\mu^4\over \mu^4}\right)
}}
In the limit $|Y|\to \mu$ (which is where the potential for $Y$ is
minimized \ISS), it reduces to
\eqn\mXrcii{
m_X^2 \to  {h^4\mu^2(\log 4-1)\over 8\pi^2}
}
which is precisely the answer obtained in \ISS.

\listrefs

\end